# Modeling L1 Influence on L2 Pronunciation: An MFCC-Based Framework for Explainable Machine Learning and Pedagogical Feedback

Peyman Jahanbin

Peyman@KU.edu

https://orcid.org/0000-0002-5692-3379

**Abstract**

This study investigates the extent to which Mel-Frequency Cepstral Coefficients (MFCCs) capture first language (L1) transfer in extended second language (L2) English speech. Speech samples from Mandarin and American English L1 speakers were extracted from the GMU Speech Accent Archive, converted to WAV format, and processed to obtain thirteen MFCCs per speaker. A multi-method analytic framework combining inferential statistics (t-tests, MANOVA, Canonical Discriminant Analysis) and machine learning (Random Forest classification) identified MFCC-1 (broadband energy), MFCC-2 (first formant region), and MFCC-5 (voicing and fricative energy) as the most discriminative features for distinguishing L1 backgrounds. A reduced-feature model using these MFCCs significantly outperformed the full-feature model, as confirmed by McNemar's test and non-overlapping confidence intervals. The findings empirically support the Perceptual Assimilation Model for L2 (PAM-L2) and the Speech Learning Model (SLM), demonstrating that L1-conditioned variation in L2 speech is both perceptually grounded and acoustically quantifiable. Methodologically, the study contributes to applied linguistics and explainable AI by proposing a transparent, data-efficient pipeline for L2 pronunciation modeling. The results also offer pedagogical implications for ESL/EFL instruction by highlighting L1-specific features that can inform intelligibility-oriented instruction, curriculum design, and speech assessment tools.

# 1. Introduction

Pronunciation plays a central role in second language (L2) communication, shaping not only intelligibility and comprehensibility but also listeners' perceptions of speaker competence (Derwing & Munro, 2005, 2009). While learners often improve in grammar and fluency, pronunciation remains a persistent barrier—especially when segmental or suprasegmental features are shaped by first language (L1) influence (Munro & Derwing, 2011). Despite growing acknowledgment of its importance, pronunciation instruction is often underrepresented in language curricula, partly due to teachers' uncertainty about which features to prioritize and how to assess them effectively (Foote et al., 2012; Murphy, 2014). Much of the feedback that learners receive is based on impressionistic judgments, which are inherently subjective and prone to rater variability (Isaacs & Thomson, 2013). This subjectivity not only affects instructional decision-making but also undermines consistency in pronunciation assessment, particularly in high-stakes contexts. In response, researchers have called for more objective, data-driven approaches to pronunciation research and instruction—especially those that illuminate systematic L1-influenced patterns in L2 speech and provide interpretable insights for teachers, learners, and automated systems alike (Derwing & Munro, 2009; Levis, 2018).

The influence of a speaker's first language (L1) on second language (L2) pronunciation is a well-established phenomenon in applied linguistics and speech sciences (Flege, 1995; Major, 2001). L1 transfer affects both segmental and suprasegmental dimensions of speech, including phonemic inventories, syllable structures, stress patterns, and intonation contours (Best & Tyler, 2007; Munro & Derwing, 2011). For instance, Mandarin Chinese speakers learning English may exhibit vowel centralization, reduced fricative precision, or deviant prosodic patterns due to differences in phonological inventories and phonotactic constraints between the two languages

(Wang & Van Heuven, 2006). While such patterns of cross-linguistic influence have been extensively studied in controlled tasks, relatively few studies have explored how L1-specific features manifest in extended or spontaneous L2 speech—despite evidence that pronunciation challenges become more pronounced under such conditions (Trofimovich & Isaacs, 2012; Derwing & Munro, 2015).

Building on recent advances in speech signal processing, researchers now have access to fine-grained, objective acoustic measures for analyzing second language (L2) pronunciation. Among these, Mel-Frequency Cepstral Coefficients (MFCCs) are particularly well-suited to capturing subtle articulatory differences, as they model short-term spectral energy distribution in a way that reflects human auditory perception (Zheng et al., 2001). MFCCs are sensitive to key phonetic features such as vowel quality, voicing, consonantal place, and manner of articulation. While they are widely used in automatic speech recognition and speaker identification, their pedagogical potential remains underutilized in applied linguistics and L2 research, where subjective ratings of intelligibility, comprehensibility, and accentedness continue to dominate pronunciation assessment (Levis, 2018, p. 214; Isaacs & Trofimovich, 2012). By leveraging MFCCs for group-level comparison, researchers can move toward more scalable and interpretable analyses of L1-influenced pronunciation—offering insights that serve both instructional and technological applications.

To address these limitations, the present study applies MFCC-based acoustic analysis in combination with inferential statistics and machine learning to examine whether specific MFCC features reliably distinguish L2 English speech produced by Mandarin and English L1 speakers. Unlike previous studies that focused on isolated segments, this research uses extended, semi-spontaneous speech to explore pronunciation patterns in more ecologically valid contexts (Wang & van Heuven, 2006). By quantifying group-level acoustic differences and identifying which MFCCs contribute most to L1-based variation, the study aims to inform both human and AI-assisted feedback systems. This work contributes to the growing body of research that bridges applied phonetics, second language acquisition, and educational technology. It also builds on earlier efforts to integrate data-informed insights across domains of language learning (Nushi & Jahanbin, 2024), supporting pedagogical innovation through quantitative modeling.

Accordingly, the study addresses the following research questions:

1. Which MFCC features show statistically significant group-level differences between Mandarin and English L1 speakers in their L2 English speech?
2. Which subset of MFCC features yields the highest classification accuracy in distinguishing between Mandarin Chinese and English L1 speakers when modeled using a Random Forest classifier?
3. To what extent do MFCC features identified through traditional inferential statistics align with those prioritized by machine learning models, and what implications does this convergence hold for L2 pronunciation assessment and feedback?

By integrating acoustic-prosodic modeling with both inferential and machine learning approaches, this study contributes to a growing movement toward data-driven and interpretable analyses of L2 pronunciation. The findings offer implications for educators, raters, and AI developers seeking to deliver more targeted, transparent, and equitable feedback to language learners.

## 2. Review of the Literature

### 2.1. Overview of L1 Influence on L2 Pronunciation

The influence of a speaker's first language (L1) on second language (L2) pronunciation is extensively documented in applied linguistics and speech sciences. Cross-linguistic transfer affects both segmental dimensions (e.g., consonants, vowels) and suprasegmental aspects (e.g., stress, rhythm, intonation), resulting in systematic deviations from native-like production (Flege, 1995; Best & Tyler, 2007). Such deviations typically arise from structural mismatches between the L1 and L2 phonological systems—including differences in phonemic inventories, syllable structure, and phonotactic constraints—and often persist even among advanced bilingual speakers (Major, 2001; Flege et al., 2003).

Mandarin Chinese speakers exemplify these challenges, frequently struggling to differentiate English tense–lax vowel contrasts such as /iː/–/ɪ/ and /uː/–/ʊ/, primarily due to cross-linguistic differences in vowel duration and spectral characteristics (Wang & van Heuven, 2006). Bent et al. (2024) demonstrated that such segmental errors significantly impair intelligibility, particularly in ecologically valid listening environments involving background noise. Similarly, Shi and Shih (2023) observed persistent difficulties among advanced Mandarin speakers in accurately producing English obstruent voicing contrasts during spontaneous speech. Listener evaluations of comprehensibility and accentedness have also been linked consistently to segmental accuracy and prosodic fluency, both strongly conditioned by speakers' L1 backgrounds (Trofimovich & Isaacs, 2012).

Theoretical models such as the Perceptual Assimilation Model for L2 (PAM-L2; Best & Tyler, 2007) and the Speech Learning Model (SLM; Flege, 1995) provide explanatory frameworks for these phenomena. PAM-L2 suggests that learners interpret new phonological input through existing L1 phonetic categories, shaping perceptual and productive challenges. Conversely, the SLM highlights learners' capacity to form distinct L2 phonetic categories if acoustic-phonetic distinctions between languages are sufficiently salient. Both models emphasize the role of detailed acoustic information—such as formant structure and spectral features—in pronunciation development.

Nevertheless, existing research predominantly employs controlled elicitation tasks, such as isolated word reading or scripted sentences, limiting generalizability to authentic communicative contexts. Such methods, while valuable for isolating specific phonetic features, do not fully capture the complexity of connected spontaneous speech, where segmental and suprasegmental cues dynamically interact. Therefore, there is an emerging need for scalable, objective methodologies that quantify L1-influenced variation in extended spontaneous speech. The present study directly addresses this gap by applying acoustic modeling approaches, using signal-based features to comprehensively analyze L1-driven pronunciation variation within authentic communicative contexts.

## 2.2. Subjectivity in Pronunciation Assessment and the Need for Acoustic Measures

Comprehensibility—defined as the listener's ease or difficulty in understanding non-native speech—has emerged as a central construct in L2 pronunciation research, alongside accentedness and fluency (Crowther et al., 2015). Typically assessed through listener judgments, these constructs are inherently subjective, influenced by raters' familiarity with specific accents, prior exposure to non-native speech, and professional background (Isaacs & Thomson, 2013; Kang et al., 2020). Isaacs and Thomson (2013), for instance, demonstrated considerable variability in raters' internal criteria and interpretive strategies, despite similar overall ratings.

Recent research has sought to identify objective linguistic features underlying these subjective evaluations. Early studies emphasized suprasegmental features: Kang (2010) identified speech rate, mean length of run, and phonation-time ratio as significant predictors of comprehensibility, whereas pitch range and word stress primarily affected accentedness. Yet growing evidence underscores segmental accuracy as a more robust predictor. Studies by Kang et al. (2020) and Saito et al. (2016) found vowel and consonant accuracy explained significantly more variance in comprehensibility and intelligibility than fluency or prosodic cues. Specifically, segmental accuracy measures operationalized through posterior probabilities exhibited superior predictive power compared to prosodic characteristics like pitch and intensity variation (Saito et al., 2023).

Despite these insights, pronunciation assessment remains heavily dependent on holistic scalar ratings. Such subjective evaluations, although practical, often obscure the precise phonetic features underlying comprehensibility, limiting diagnostic utility, especially in extended spontaneous speech contexts (Isbell, 2020; Isaacs & Thomson, 2013). Hence, there is a clear need for objective acoustic measures that pinpoint phonetic accuracy.

Spectral descriptors—particularly Mel-Frequency Cepstral Coefficients (MFCCs)—represent a promising yet underutilized approach. MFCCs quantify short-term spectral dynamics directly related to articulatory and phonetic features. The current study addresses this critical gap by employing MFCC-based acoustic modeling to spontaneous English speech produced by

Mandarin Chinese and English L1 speakers, establishing a precise and data-driven framework for pronunciation assessment.

## 2.3. Integrating Spectral Features into Acoustic Modeling of L1 Influence

Acoustic-prosodic research in L2 pronunciation has traditionally prioritized suprasegmental features such as speech rate, pitch, and pausing patterns, often neglecting spectral cues (Bent et al., 2024; Li & Post, 2014). For example, Bent et al. (2024) found that pause structures and syllable reductions were key predictors of comprehensibility in extended L2 speech. Similarly, Li and Post (2014) demonstrated that rhythmic-temporal measures—including vocalic variability and accentual lengthening—varied systematically across L1 backgrounds, providing quantitative evidence of prosodic transfer in L2 English. However, neither study incorporated spectral parameters such as Mel-Frequency Cepstral Coefficients (MFCCs), despite their sensitivity to articulatory features like vowel quality, voicing, and place of articulation (Zheng et al., 2001).

MFCCs, though common in speech recognition and engineering, remain underrepresented in applied linguistics. Existing studies that utilize MFCCs have predominantly been limited to controlled, laboratory-based phoneme analyses, significantly restricting ecological validity and practical applicability (Wang & van Heuven, 2006). Wang and van Heuven (2006), for instance, examined vowel production using isolated syllables, limiting the relevance to connected, spontaneous speech contexts. Moreover, research typically employs either inferential statistical methods or machine learning approaches exclusively, rarely combining both to simultaneously identify statistically significant and computationally robust spectral features.

Nevertheless, some studies underscore MFCCs' potential for modeling L1-driven pronunciation variations effectively. Shi and Shih (2019) employed MFCC-derived acoustic distance metrics alongside PCA to predict systematic vowel deviations in Mandarin-accented English, validating predictions from PAM-L2 and SLM frameworks. Still, their work focused solely on isolated vowels, leaving open the question of applicability to spontaneous speech.

The present study directly addresses these methodological gaps by applying all 13 standard MFCCs to spontaneous speech data from Mandarin Chinese and English L1 speakers. Integrating inferential statistics and advanced machine learning classification (Random Forest), this study identifies MFCC features that most clearly differentiate L1 backgrounds, contributing to more interpretable, data-driven acoustic modeling with instructional and assessment implications.

**2.4. Acoustic-Prosodic Modeling: Towards Integrated Methodologies**

Advancements in automated speech processing have broadened opportunities to objectively analyze L2 pronunciation through integrated acoustic-prosodic modeling. Recent computational approaches increasingly combine spectral features (e.g., MFCCs) with suprasegmental parameters such as pitch and duration, providing scalable alternatives to subjective listener-based assessments (Saito et al., 2023; Shivakumar et al., 2016).

Initial studies like Patil et al. (2013) demonstrated MFCCs' effectiveness at capturing phoneme-level pronunciation deviations through Dynamic Time Warping (DTW). Shivakumar et al. (2016) extended these findings, employing neural network models with MFCCs and low-level descriptors (LLDs) to predict accentedness ratings. More advanced computational frameworks such as Chao et al.'s (2022) 3M model have integrated MFCCs into multi-view architectures, successfully assessing pronunciation across multiple linguistic granularities (phoneme, syllable, word).

Parallel linguistic research typically emphasizes prosodic parameters, highlighting persistent L1 influence in rhythm and pitch, even among proficient learners (Silva Jr., & Meer, 2024). Yet, spectral measures remain comparatively underexplored within spontaneous speech contexts. This study addresses this imbalance by employing comprehensive MFCC-based acoustic modeling on spontaneous speech from Mandarin and English L1 speakers, combining inferential statistical analyses and Random Forest classification to develop robust, interpretable acoustic models. By integrating segmental and suprasegmental insights, this unified methodology significantly advances acoustic-prosodic modeling, informing practical applications in pronunciation pedagogy and automated assessment.

## 3. Methodology

### 3.1. Data Source and Selection

The speech samples analyzed in this study were obtained from the George Mason University Speech Accent Archive (Weinberger, 2015), a publicly accessible corpus that includes standardized readings of a scripted paragraph by speakers from a variety of first language (L1) backgrounds. A total of 58 recordings were selected from native speakers of American English originating in the Midwest region of the United States, alongside 60 recordings from native speakers of Mandarin Chinese from mainland China. All recordings featured the same scripted paragraph, designed to elicit a wide range of English phonemes in connected speech. The audio files were downloaded in MP3 format and converted to 44.1 kHz mono WAV files for subsequent processing.

**3.2. Acoustic Feature Extraction**

The acoustic analysis was conducted in Python using the Librosa library within the Google Colab environment. Figure 1 illustrates a comprehensive overview of the entire analysis pipeline. Thirteen Mel-Frequency Cepstral Coefficients (MFCCs) were extracted for each recording using a 25-millisecond analysis window with a 10-millisecond hop size, standard parameters in speech processing (Rabiner & Schafer, 2011). MFCCs are widely used in speech analysis due to their ability to capture phonetically relevant spectral features (Davis & Mermelstein, 1980; Zheng et al., 2001). To obtain a single representative value for each feature, the mean value of each MFCC across all frames in a speaker's recording was calculated, resulting in a 13-dimensional feature vector per speaker. All acoustic preprocessing and feature extraction were performed using Python 3.10 in the Google Colab cloud-based environment. The Librosa library (v0.10.0) was used for MFCC extraction, and NumPy and Pandas were employed for data manipulation.

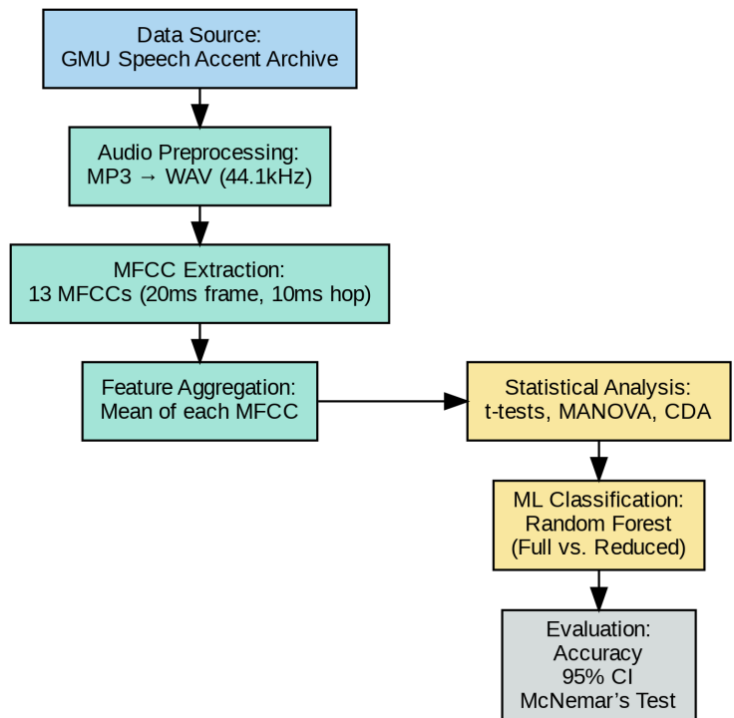

**Figure 1**
*MFCC-based analysis pipeline for L2 pronunciation modeling, integrating feature extraction, statistical testing, and classification.*

### 3.3. Statistical Analysis

To assess whether individual MFCCs differed significantly between the two language groups, independent samples t-tests were conducted for each feature, with Bonferroni correction applied to adjust for multiple comparisons. A one-way multivariate analysis of variance (MANOVA) was then used to examine group differences across the full MFCC vector. Assumptions of multivariate normality and homogeneity of variance-covariance matrices were evaluated, and Pillai's Trace was used as the omnibus test statistic. Canonical Discriminant Analysis (CDA) was subsequently employed to determine which MFCCs contributed most to group separation, based on standardized canonical coefficients (Tabachnick & Fidell, 2019).

### 3.4. Machine Learning Classification

A Random Forest classifier (Breiman, 2001) was implemented using the scikit-learn library (Pedregosa et al., 2011) to assess the discriminative power of the extracted MFCC features. Two models were trained: one using the complete set of 13 MFCCs and the other utilizing a reduced subset consisting of MFCC_1, MFCC_2, and MFCC_5. These features were chosen based on their statistical significance in independent-samples t-tests (with Bonferroni correction) and their high standardized canonical coefficients in Canonical Discriminant Analysis (CDA), highlighting their significant contribution to group separation. Model performance was assessed using classification accuracy and 95% confidence intervals. To test the significance of the difference in classification performance between the two models, McNemar's test was applied (Dietterich, 1998). Confidence intervals were computed using the Wilson score method for binomial proportions (Brown et al., 2001). All machine learning procedures were implemented in Python using the scikit-learn library (v1.2.2) (Pedregosa et al., 2011) within the same Google Colab environment.

### 3.5 Open Science and Reproducibility

To support transparency and reproducibility, the complete analysis pipeline—including audio preprocessing, MFCC extraction, statistical modeling, and classifier implementation—has been archived on Zenodo (Jahanbin, 2025a) and permanently registered on the Open Science Framework (Jahanbin, 2025b). Shared materials include the MFCC extraction script, SPSS output file, and full documentation. All analyses were conducted in Python (v3.10) using Librosa and scikit-learn, and statistical tests were performed in SPSS. These resources offer a reproducible foundation for future work in L2 pronunciation modeling, instructional design, and AI-assisted assessment.

## 4. Data Analysis

### 4.1. Descriptive Statistics

Descriptive statistics were computed to summarize the central tendency and dispersion of the 13 Mel-Frequency Cepstral Coefficients (MFCC1–MFCC13) for the two language groups: Mandarin Chinese ($n = 60$) and American English ($n = 58$). For each MFCC, the mean ($M$) and standard deviation ($SD$) were calculated. These descriptive metrics offer a quantitative profile of the acoustic characteristics of each group based on a standardized 40-second paragraph reading. They serve as the foundation for subsequent statistical comparisons, enabling controlled group-level analysis of L2 speech production. Table 1 presents the full descriptive statistics for all 13 MFCCs across both language groups.

### 4.2. Inferential Statistics: Independent Samples t-test

#### 4.2.1 Assumption Checks

Tests of normality were conducted using the Shapiro–Wilk and Kolmogorov–Smirnov statistics. Although a few MFCCs exhibited minor deviations from normality (e.g., MFCC13 for English speakers, $p = .046$), the majority of distributions did not significantly deviate ($p > .05$), supporting the use of parametric procedures. Levene's test for equality of variances indicated that the assumption of homogeneity was met for most coefficients (e.g., MFCC1, $p = .213$). For coefficients where this assumption was violated (e.g., MFCC3, $p = .015$; MFCC5, $p = .028$), Welch's *t*-test was employed to account for heteroscedasticity.

### 4.2.2 Group Comparisons

Independent samples *t*-tests revealed statistically significant differences between groups on several MFCCs, including MFCC1 (broadband energy), MFCC2 (often associated with spectral slope or lower formant regions), MFCC5 (potentially capturing fricative-related energy), and MFCC9 (higher-frequency dynamics), all at the conventional alpha level of .05 ($p < .05$), as shown in Table 1. While these interpretations are heuristic, they are consistent with foundational and empirical analyses of MFCC functionality (Davis & Mermelstein, 1980; Zheng, Zhang, & Song, 2001).

To control for Type I error across 13 comparisons, a Bonferroni correction was applied, adjusting the significance threshold to $\alpha = .0038$. After this correction, only MFCC5 remained statistically significant ($p = .002$, $d = 0.541$), suggesting that it is the most robust and reliable indicator of pronunciation differences between the two groups. A boxplot of MFCC5 illustrates this group-level contrast and visualizes the distributional differences between English and Mandarin speakers (see Figure 2). While other coefficients showed potentially meaningful trends, they did not reach significance under the more stringent criterion. Additional boxplots for

MFCC1, MFCC2, and MFCC9 are provided in Figure 3 to illustrate group-level patterns that did not remain statistically significant after the Bonferroni correction.

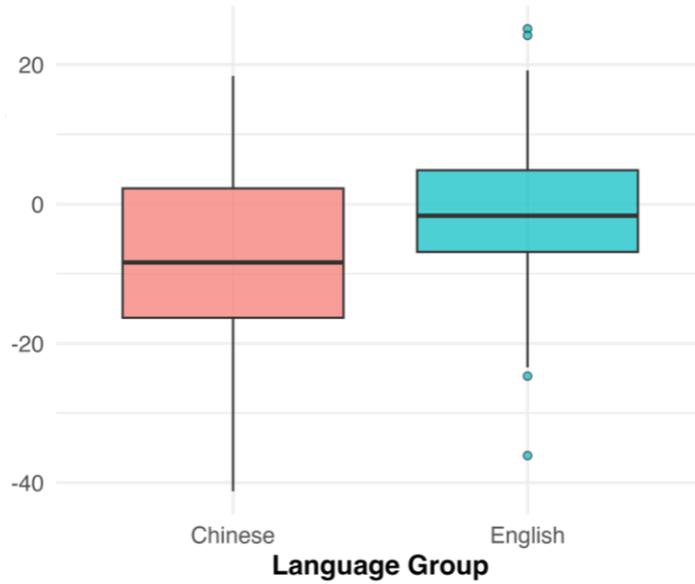

**Figure 2**
*Boxplot of MFCC5 (Onset Frictional Voicing Features) for English and Mandarin Speakers*

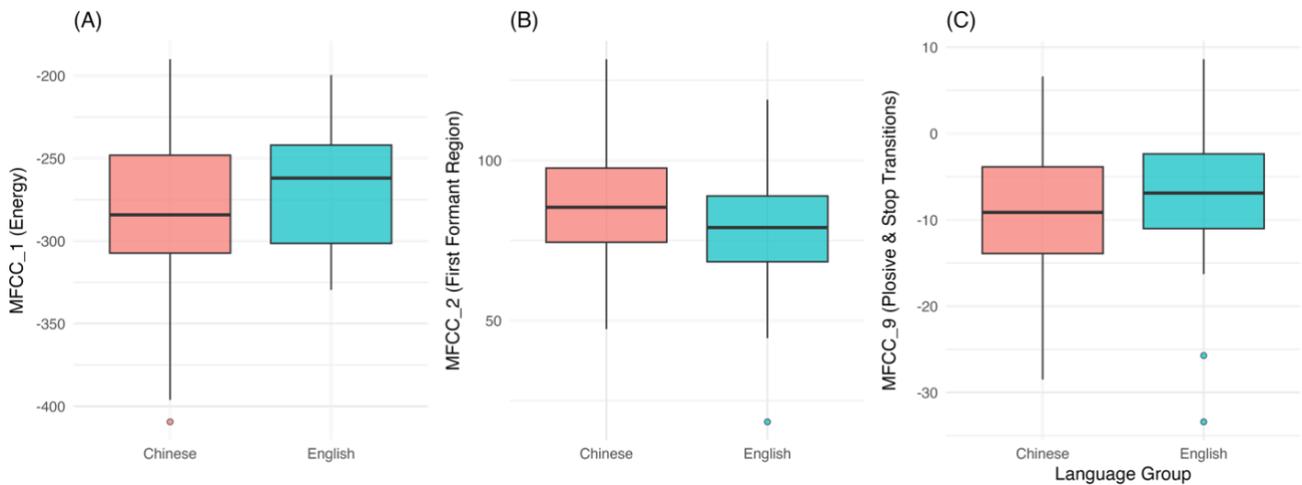

**Figure 3**
*Boxplots of MFCC1 (Energy), MFCC2 (First Formant Region), and MFCC9 (Plosive and Stop Transitions)*

| MFCC feature | English (*M* ± *SD*) | Mandarin (*M* ± *SD*) | *t*(df) | *p* | 95% CI of Mean Diff. | Cohen's d |
|---|---|---|---|---|---|---|
| MFCC1_Energy | -266.87 ± 34.56 | -282.12 ± 44.02 | -2.09 (116) | .039 | [-29.72, -0.80] | 0.385 |
| MFCC2_First Formant Region | 79.23 ± 18.78 | 86.71 ± 19.78 | 2.10 (116) | .038 | [0.44, 14.51] | 0.387 |
| MFCC3_Mid Formant Region | 3.61 ± 12.22 | 3.58 ± 15.75 | -0.01 (116) | .990 | [-5.19, 5.12] | -0.002 |
| MFCC4_Voicing & Nasal Characteristics | 24.51 ± 10.74 | 22.65 ± 13.50 | -0.83 (116) | .411 | [-6.31, 2.60] | -0.152 |
| MFCC5_Onset Frictional Voicing | -0.99 ± 10.92 | -7.69 ± 13.63 | -2.95 (112.19) | .002 | [-11.19, -2.20] | 0.541 |
| MFCC6_Coarticulation & Vowel Rounding | -3.83 ± 11.02 | -3.64 ± 11.22 | 0.09 (116) | .925 | [-3.86, 4.25] | 0.017 |
| MFCC7_Spectral Tilt | -9.74 ± 9.93 | -11.61 ± 9.77 | -1.03 (116) | .306 | [-5.46, 1.73] | -0.189 |
| MFCC8_Aspiration/Voicing Transitions | -3.10 ± 7.80 | -5.22 ± 11.31 | -1.19 (105.01) | .236 | [-5.66, 1.41] | -0.218 |
| MFCC9_Plosive & Stop Transitions | -6.86 ± 7.45 | -9.69 ± 7.89 | -2.00 (116) | .048 | [-5.63, -0.03] | 0.368 |
| MFCC10_Second Formant-Like Shape | -2.83 ± 6.71 | -4.57 ± 7.30 | -1.35 (115.72) | .179 | [-4.30, 0.81] | -0.248 |
| MFCC11_Back Vowel Region | -4.95 ± 5.55 | -6.70 ± 6.98 | -1.50 (116) | .136 | [-4.05, 0.56] | -0.277 |
| MFCC12_Intonation or Pitch Influences | -1.81 ± 5.56 | -2.72 ± 6.52 | -1.71 (114.24) | .089 | [-4.12, 0.30] | -0.315 |
| MFCC13_Higher-order Subtle Cues | -3.52 ± 5.05 | -4.43 ± 5.87 | -0.90 (114.49) | .370 | [-2.90, 1.09] | -0.166 |

**Table 1**

*Independent Samples t-Test Results for MFCCs by Language Group*

*Note*. Welch's *t*-test was used where Levene's test indicated unequal variances. Bonferroni correction applied: adjusted *α* = .0038.

### 4.2.3 Multivariate Analysis of Variance (MANOVA)

A one-way multivariate analysis of variance (MANOVA) was conducted to assess whether the combined MFCCs 1–13 differed significantly across the two language groups (English and Mandarin) (Tabachnick & Fidell, 2019). Prior to the analysis, Box's M test was examined to assess the assumption of homogeneity of covariance matrices. The test was significant, *Box's M* = 144.02, *p* = .008, suggesting a violation of this assumption. Accordingly, the results are interpreted using Pillai's Trace, which is considered more robust to such violations (Olson, 1974; Tabachnick & Fidell, 2019).

The overall MANOVA was statistically significant, *Pillai's Trace* = 0.192, $F(13, 104)$ = 1.90, $p$ = .038, indicating a multivariate effect of language group on the combined MFCC features. This corresponds to a partial $\eta^2$ of 0.24, indicating that approximately 24% of the variance in the combined acoustic feature space can be attributed to language group membership—a small-to-moderate effect by conventional thresholds (Cohen, 1988; Tabachnick & Fidell, 2019).

**4.2.4 Canonical Discriminant Analysis (CDA)**

Following the significant MANOVA, a Canonical Discriminant Analysis (CDA) was conducted to further examine group separation. As only two language groups were compared, the analysis produced a single canonical function (Can1), which accounted for the total observed between-group variance. The standardized canonical coefficients indicated that MFCC5 (onset frictional voicing features), MFCC1 (energy), and MFCC2 (first formant region) were the primary contributors to the group discrimination (Tabachnick & Fidell, 2019). As shown in Figure 4, the canonical scores plot demonstrates a clear separation between English and Mandarin speakers along Can1, reinforcing the discriminant power of the MFCCs. While other MFCCs (e.g., MFCC9) contributed to the canonical function, their loadings were relatively smaller and consistent with the univariate results that did not remain significant after the Bonferroni correction.

Together, these findings suggest that a constellation of acoustic features, especially those related to voicing dynamics and spectral energy distribution, distinguishes L1 pronunciation patterns in English and Mandarin speakers. The CDA confirms and extends the t-test findings by modeling the MFCCs jointly and revealing their combined discriminatory strength.

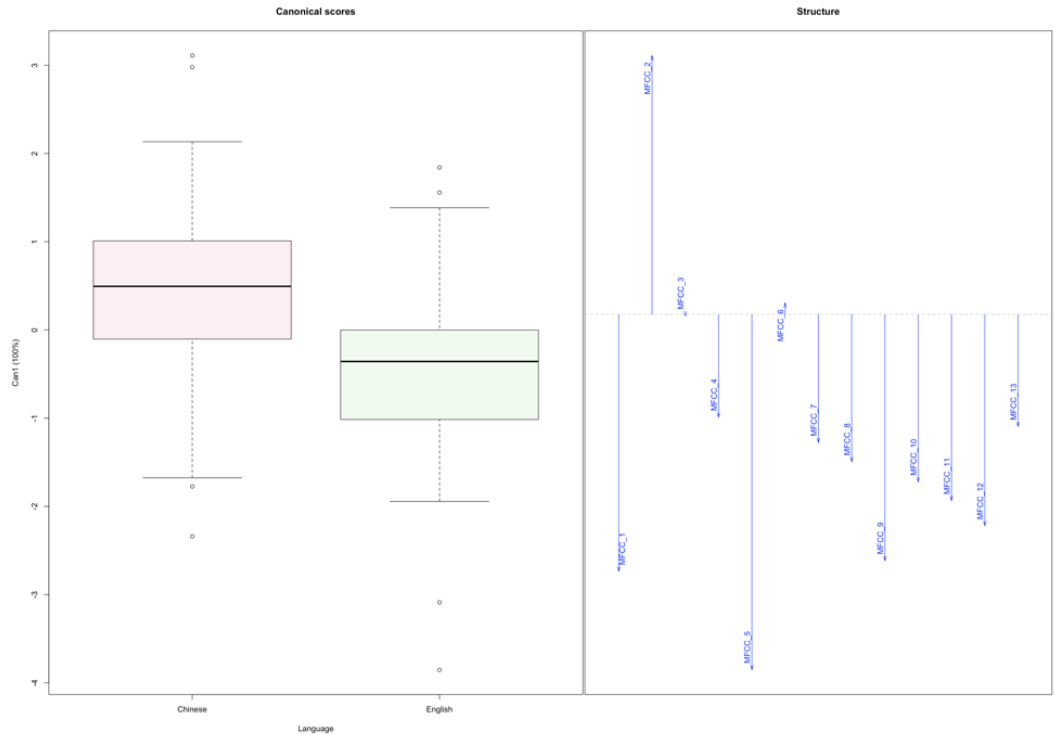

**Figure 4**
*Canonical Discriminant Analysis of MFCC Features by Language Group*

### 4.2.5 Random Forest Classifier

To complement the inferential analysis and assess the practical discriminative utility of the MFCC features, a Random Forest classifier was trained using only the MFCCs that demonstrated both statistical significance in prior analyses and strong contributions in the Canonical Discriminant Analysis—specifically, MFCC_1 (broadband energy), MFCC_2 (first formant region), and MFCC_5 (onset frictional voicing features). This reduced-feature model achieved a classification accuracy of 75.00% (95% CI: [58.93%, 86.25%] Wilson method), notably higher than the 52.78% accuracy (95% CI: [37.01%, 68.01%]) observed when all 13 MFCCs were included. McNemar's test (Dietterich, 1998) confirmed that the difference in performance between the two models was statistically significant ($p = .039$), suggesting that the improved accuracy of the reduced model is unlikely due to chance alone. The Random Forest classifier, chosen for its ability to handle high-dimensional data and rank feature importance effectively (Breiman, 2001), further validated the practical relevance of the selected MFCC features.

The decision to exclude MFCC_9, despite its initial significance in the t-tests, was guided by its comparatively weaker canonical loading in the CDA, suggesting limited multivariate contribution to group separation. The superior performance of the reduced model underscores the importance of data-driven feature refinement—where only those features supported by both univariate and multivariate evidence are retained for modeling. This approach aligns with best practices in multivariate statistics (Tabachnick & Fidell, 2019) and feature selection research (Guyon et al., 2003). Moreover, the use of confidence intervals provides a more transparent and interpretable representation of classifier reliability, reinforcing the reduced model's empirical advantage.

This convergence between traditional statistical inference and machine-learned feature importance highlights the interpretability and practical relevance of MFCC_1, MFCC_2, and MFCC_5 in distinguishing first language (L1) backgrounds from extended second language (L2) speech samples. The results offer implications not only for pronunciation instruction but also for automated speech classification, enabling more efficient feedback systems and improved L1 detection based on theoretically motivated acoustic indicators. Together, the statistical, practical, and interpretive significance of the reduced-feature model supports its use in future applications of second language pronunciation modeling and assessment.

## 5. Discussion and Pedagogical Implications

The present study investigated how Mel-Frequency Cepstral Coefficients (MFCCs) can capture first language (L1)–induced variation in extended second language (L2) English speech.

Through a triangulated analysis using inferential statistics, multivariate modeling, and machine learning classification, the study identified MFCC_1 (broadband energy), MFCC_2 (first formant region), and MFCC_5 (onset voicing and fricative energy) as the most reliable acoustic features for distinguishing Mandarin and English L1 speakers. These findings were consistently supported across t-tests, MANOVA, Canonical Discriminant Analysis (CDA), and Random Forest classification, offering a robust and interpretable profile of L1-conditioned pronunciation features with implications for both learner-centered instruction and automated classification systems.

The integration of McNemar's test (Dietterich, 1998) and confidence intervals further confirmed the practical value of these features: a reduced-feature model using only MFCC_1, MFCC_2, and MFCC_5 achieved significantly higher classification accuracy (75.00%) compared to the full 13-feature model (52.78%). This result highlights the effectiveness of data-driven feature refinement guided by theoretical and statistical insight. Importantly, this performance gain was not only statistically significant (p = .039) but also supported by non-overlapping confidence intervals, underscoring both the practical and empirical value of targeted acoustic modeling.

These results align closely with the predictions of established second language speech theories. The Perceptual Assimilation Model for L2 (PAM-L2; Best & Tyler, 2007) posits that L2 learners perceive novel phonemes through the lens of their L1 phonological system, and that difficulty arises when L2 contrasts are assimilated into a single L1 category. MFCC_1 and MFCC_2, which capture energy and vowel formant structure, directly reflect such perceptual categories and their potential assimilation. Similarly, the Speech Learning Model (SLM; Flege, 1995) argues that learners may form new phonetic categories if they can perceive sufficient acoustic distance between L1 and L2 realizations. The strong discriminability of MFCC_5—which relates to voicing and fricative features—suggests that certain consonantal cues may indeed be perceptually distinct enough to support category formation. Thus, the acoustic differences captured by MFCCs align with the phonetic cues identified as central in PAM-L2 and SLM, confirming that L1-induced variation is both theoretically predictable and empirically measurable.

Beyond its theoretical implications, the study offers practical insights for pronunciation instruction and language assessment. The identification of a small, high-performing set of MFCCs provides a foundation for developing targeted diagnostic tools that highlight L1-specific pronunciation targets in learner speech, such as vowel reduction and fricative devoicing, to guide individualized feedback. For example, MFCC_1 reflects overall vowel energy and may signal reduced articulatory effort or vowel reduction in L2 speech, while MFCC_2 corresponds to the first formant region, often associated with tongue height and frontness—key areas of difficulty for many Mandarin-speaking learners of English. MFCC_5 captures voicing and fricative energy, potentially reflecting difficulties in producing voiced fricatives and stops accurately. Awareness of such acoustic markers can support English language teachers, curriculum designers, and language test developers in prioritizing pronunciation targets that are both perceptually salient and phonologically motivated.

English language teachers working with Mandarin L1 learners could incorporate targeted pronunciation activities that explicitly focus on vowel contrasts (e.g., /iː/ vs. /ɪ/) and voiced fricatives (e.g., /v/, /z/), which are frequently affected by L1 transfer. Instructional strategies might include minimal pair discrimination, delayed repetition, and guided shadowing activities designed to increase learners' auditory sensitivity to these contrasts. Additionally, teachers could draw on simplified visual supports—such as vowel space maps or voiced/unvoiced contrast charts—to reinforce articulatory explanations and support learner noticing. Curriculum designers might embed such contrastive tasks into pronunciation modules tailored to L1-specific patterns, promoting more efficient and targeted pronunciation development.

For language test developers, the MFCC-based acoustic profiles identified in this study could inform the development of analytic rating scales by emphasizing intelligibility-relevant features such as vowel clarity and voicing accuracy. These data-driven insights could enhance the construct validity of oral assessments and support more diagnostic approaches to scoring. Furthermore, MFCC-informed acoustic modeling can be integrated into AI-assisted pronunciation tutors, enabling real-time, individualized feedback on pronunciation errors that are both phonetically significant and pedagogically interpretable. This alignment between acoustic evidence and instructional practice offers a scalable pathway for developing tools that are not only technologically sophisticated but also educationally meaningful.

This study contributes methodologically by demonstrating how triangulation across traditional statistical inference and machine learning can enhance both the rigor and practical relevance of L2 speech research. Rather than treating statistical and computational methods as separate silos, this study shows that integrating them can yield insights that are theoretically informed, empirically grounded, and pedagogically actionable.

From a technological perspective, this study contributes to the development of interpretable and data-efficient AI systems for speech classification. The reduced-feature Random Forest model demonstrated that three well-selected MFCCs could outperform a full 13-feature model, highlighting the value of feature selection for model compression and deployment in real-world applications. Moreover, by combining inferential statistics, discriminant analysis, and transparent classification models, the study exemplifies an explainable AI (XAI) approach to pronunciation modeling (Doshi-Velez & Kim, 2017)—supporting the development of educational tools that are both effective and pedagogically interpretable. These insights are particularly relevant for designers of AI-assisted pronunciation tutors and L2 assessment tools, where system transparency and alignment with linguistic theory are increasingly prioritized. Finally, the identification of cross-linguistically salient MFCC features offers a scalable foundation for multilingual speech recognition systems and adaptive feedback platforms that dynamically tailor instruction to learners' L1-specific pronunciation profiles—enhancing both personalization and effectiveness.

Future research could extend this approach to other L1 groups and examine whether MFCC-based classification aligns with human listeners' judgments of comprehensibility and accentedness in longitudinal classroom contexts. Such extensions would further validate the pedagogical relevance of acoustic-prosodic modeling in second-language pronunciation instruction. Future research may also explore integrating MFCC-based models into classroom-based formative assessment systems or pronunciation tutoring platforms to assess their effectiveness in improving learner outcomes over time.

## 6. Conclusion

This study investigated the extent to which Mel-Frequency Cepstral Coefficients (MFCCs) capture first language (L1)-induced variation in extended second language (L2) English speech, specifically comparing Mandarin Chinese and native English speakers. Through a multi-method approach—integrating inferential statistics, Canonical Discriminant Analysis, Random Forest classification, and McNemar's test—the analysis identified MFCC_1, MFCC_2, and MFCC_5 as the most discriminative and theoretically interpretable features. These features not only improved classification accuracy but also aligned with key tenets of the Perceptual Assimilation Model for L2 (PAM-L2) and the Speech Learning Model (SLM), reinforcing the claim that L1-conditioned variation is both acoustically measurable and perceptually grounded.

Beyond its empirical findings, the study contributes to the development of interpretable, data-efficient, and pedagogically actionable models for L2 pronunciation research. By isolating a small subset of MFCCs that are both statistically robust and linguistically meaningful, the study informs the design of AI-assisted feedback systems that can be aligned with classroom instruction and curricular priorities. Moreover, it illustrates the methodological value of combining statistical inference with explainable machine learning to advance both applied linguistics and speech technology research.

Future research should extend this framework by incorporating suprasegmental features, examining additional L1 groups, and analyzing learner speech across proficiency levels. Such work would further clarify the dynamics of cross-linguistic influence and enable the development of adaptive, linguistically principled pronunciation assessment systems that respond to the needs of diverse multilingual learners in real-world instructional contexts.